# A versatile multicomponent mesoporous silica nanosystem with dual antimicrobial and osteogenic effects

*Elena Álvarez,[a,b,†] Manuel Estévez,[a,†] Carla Jiménez-Jiménez,[b] Montserrat Colilla,[a,b] Isabel Izquierdo-Barba,[a,b] Blanca González[a,b]\* and María Vallet-Regí[a,b]\**

[a] Dpto. Química en Ciencias Farmacéuticas, Universidad Complutense de Madrid, Instituto de Investigación Sanitaria, Hospital 12 de Octubre i+12, Plaza Ramón y Cajal s/n, 28040 Madrid, Spain

[b] CIBER de Bioingeniería, Biomateriales y Nanomedicina, CIBER-BBN, Madrid, Spain

[†] Both authors have contributed equally

\* Corresponding authors: blancaortiz@ucm.es; vallet@ucm.es

**ABSTRACT**

In this manuscript, we propose a simple and versatile methodology to design nanosystems based on biocompatible and multicomponent mesoporous silica nanoparticles (MSNs) for infection management. This strategy relies on the combination of antibiotic molecules and antimicrobial metal ions into the same nanosystem, affording a significant improvement of the antibiofilm effect compared to that of nanosystems carrying only one of these agents. The multicomponent nanosystem is based on MSNs externally functionalized with a polyamine dendrimer (MSN-G3) that favors internalization inside the bacteria and allows the complexation of multiactive metal ions (MSN-G3-$M^{n+}$). Importantly, the selection of both the antibiotic and the cation may be done depending on clinical needs. Herein, levofloxacin and $Zn^{2+}$ ion, chosen owing to both its antimicrobial and osteogenic capability, have been incorporated. This dual biological role of $Zn^{2+}$ could have and adjuvant effect thought destroying the biofilm in combination with the antibiotic as well as aid to the repair and regeneration of lost bone tissue associated to osteolysis during infection process. The versatility of the nanosystem has been demonstrated incorporating $Ag^+$ ions in a reference nanosystem. *In vitro* antimicrobial assays in planktonic and biofilm state show a high antimicrobial efficacy due to the combined action of levofloxacin and $Zn^{2+}$, achieving an antimicrobial efficacy above 99% compared to the MSNs containing only one of the microbicide agents. *In vitro* cell cultures with MC3T3-E1 preosteoblasts reveal the osteogenic capability of the nanosystem, showing a positive effect on osteoblastic differentiation while preserving the cell viability.

# 1. INTRODUCTION

The main cause of chronic implant-related infections is the biofilm formation [1]. It begins with the bacteria adhesion preferentially in the surface of the foreign body material and subsequently colonization, establishing a community of microorganisms embedded in a self-produced polysaccharide matrix denoted as biofilm [2]. These biofilms provide protection against external agents like antibiotics and the host's immune system [3]. In general, the presence of biofilms leads to the progress of multidrug-resistant bacteria, reappearance of infection and treatment failure [4,5]. When infection reaches the bone, biofilm provokes, in addition, a local inflammatory response resulting in stimulation of osteoclastogenesis, with local osteolysis and bone loss leading to septic loosening of the implant [6]. In general, antibiofilm therapies are not fully satisfactory, since they are mainly based on the massive systemic administration of antibiotic cocktails, usually ineffective and with serious side effects for the patient. During decades, the scientific efforts were addressed to design bulk bioactive materials, which at the same time as regenerating bone were able to eliminate the infection by incorporating antibiotics. Although these offer numerous advantages, they lack specificity and adequate release control required [7-10]. In the present, research attempts are aimed at the prevention strategies, especially the modification of implant surfaces [11-14]. However, once bacterial biofilm is established, it is necessary to eradicate it with advanced and effective therapies. On the other hand, metal ions such are re-emerging in the face of antibiotic ineffectiveness [15]. The antimicrobial efficacy of certain metal ions is very powerful against a wide variety of bacteria, without developing any resistance [16]. However, the high cytotoxicity greatly limits its wide application in clinical medicine [17]. Hence, nanotechnology has emerged as promising solution for developing effective antibacterial-based devices due to their abundant surface chemistry and high surface-to-volume ratio, accounting for ease of surface functionalization for immobilizing targeting ligands as well as encapsulating different therapeutic molecules, respectively [18,19]. In this way several nanocarriers such as liposomes, polymeric-nanodevices, and different inorganic nanoparticles have been formulated for delivery different antimicrobial agents [20,21]. In terms of stability and surface properties, inorganic nanosystems as mesoporous silica nanoparticles (MSNs) have gathered of particular interest owing to their advantageous morphological features and attractive physicochemical properties [22-24]. These nanocarriers exhibit high versatility and biocompatibility, which allows its application in a wide range of pathologies included infection treatment [25]. Their mesoporous structure allows to host several antimicrobial agents of different nature and to release them in a sustained manner over time at local level [26]. In addition, their surface allows its easy and tuneable modification obtaining nanosystems with high specificity and stimulus response release [27-29]. Previously, our research group developed MSNs able to interact with the Gram-negative bacteria membrane and to internalise inside this kind of pathogen [30]. This internalizing capability is provided by the external functionalization of the antibiotic-loaded MSNs with a polycationic dendrimer, concretely the poly(propyleneimine) dendrimer of third generation (G3). The improvement in this design is the ability to be internalised, therefore substantially improving the antimicrobial effect of the antibiotic loaded against bacteria either in a planktonic state or forming a biofilm. These achievements inspired us the idea of exploring the possibility to provide a solution to the growing problem of multi-resistant bacteria. To this aim, we propose the combination of antimicrobial metal ions and antibiotics in a unique MSN nanosystem capable of reducing biofilm and preventing the emergence of antimicrobial resistance due to the presence of these metal cations. In addition, certain metal ions exhibit dual effect that could have an adjuvant effect in the infection management. This is the case of $Zn^{2+}$, which in addition to produce a powerful antimicrobial effect it also promotes osteogenesis, thus obtaining an adjuvant effect in repairing the bone area damaged by the osteolysis process. Herein, a synthetic procedure of multicomponent antimicrobial nanosystems is described. The simple and versatile procedure developed in this work allows the incorporation of two antimicrobial agents, antibiotics and cations, in the same MSN based nanoplatform depending on clinical needs. In this case, MSNs have been loaded with LEVO a broad-spectrum antibiotic for general use in bone infection.

Likewise, $Zn^{2+}$ cations have been also incorporated due to its antimicrobial effect as well as osteogenic capability [31-34]. To demonstrate the versatile capability of such methodology, nanosystems incorporating $Ag^+$ ions have been also prepared as a reference due to the well-known antimicrobial power of silver [35]. The proposed design affords the preparation of nanosystems with combined effect of two microbicide agents possessing as well a dual effect in bone cells differentiation.

## 2. EXPERIMENTAL SECTION

### 2.1. Reagents

Fluorescein isothiocyanate (FITC), tetraethylorthosilicate (TEOS), cetyltrimethylammonium bromide (CTAB), levofloxacin (LEVO, see Fig. S1), silver nitrate ≥99.8% and zinc nitrate ≥99.8% were purchased from Sigma-Aldrich. 3-Aminopropyltriethoxysilane 97% (APTS) and 3-isocyanatopropyltriethoxysilane 95% were purchased from ABCR GmbH & Co. KG., phosphate-buffered saline (PBS) was purchased from GIBCO and the G3-PPI dendrimer [G3$(NH_2)_{16}$] from SyMO-Chem. Deionized water was further purified by passage through a Milli-Q Advantage A-10 purification system (Millipore Corporation) to a final resistivity of 18.2 MΩ cm. All other chemicals (ammonium nitrate, absolute ethanol, sodium hydroxide, etc.) were of the highest quality commercially available and used as received.

The synthesis of the silylated dendrimer G3-Si(OEt)$_3$ and the chemical functionalization of the silica surface to afford MSN-G3 were carried out following our previous reported procedure [30]. These compounds were synthesized under an inert atmosphere using Schlenk techniques. Dichloromethane was dried by standard procedures over phosphorus (V) oxide and distilled immediately prior to use. All manipulations of light sensitive compounds (fluorescein or silver containing materials) were performed with protection from light. Synthetized and antibiotic loaded nanosystems were kept refrigerated at 4 °C in dry conditions.

The analytical methods used to characterize the synthesized compounds were as follows: low-angle powder X-ray diffraction (XRD), thermogravimetry and chemical microanalysis, Fourier transformed infrared (FTIR) and UV-visible spectroscopies, transmission electron microscopy (TEM), energy dispersive X-ray spectroscopy (EDS), electrophoretic mobility measurements to calculate the values of zeta-potential (ζ), dynamic light scattering (DLS), inductively coupled plasma atomic emission spectroscopy (ICP-AES), confocal microscopy and flow cytometry. The equipment and conditions used are described in the Supporting Information.

Reagents for *in vitro* microbiological and cell assays were as follows: Luria-Bertani broth (LB), Todd Hewitt broth (THB), Tryptic Soy Agar (TSA), poly-L-lysine, sucrose, Calcofluor White Stain, Thiazolyl Blue Tetrazolium Bromide (M5655), Trypan Blue (T8154), β-glycerophosphate (G9422), ascorbic acid (A9290-2), Alizarin Red S staining (130-22-3) and cetylpyridinium chloride (C0732) from Sigma-Aldrich); α-MEM medium (22561-021) and fetal bovine serum (FBS, 10270-106) from Gibco; Live/Dead Bacterial Viability Kit from Backlight™, and penicillin/streptomycin (13-0050) from ZellShield.

## 2.2. Materials synthesis

**MSN-G3-$M^{n+}$.** The same procedure was used for the complexation of both cations using 3.5 equivalents of metal nitrate per mol of G3. Briefly, 10 mg of MSN-G3 were stirred in 5 mL of water until a homogeneous suspension was obtained. Ultrasound was also applied if needed. Then, 38.5 μL of a 0.1 M $M(NO_3)_n$ ($M^{n+}$ =$Ag^+$ or $Zn^{2+}$) solution was added dropwise and the mixture was kept under stirring for 15 minutes. Subsequently, the material was recovered via centrifugation, washed with water and EtOH and finally dried. MSN-G3-$M^{n+}$ materials were also prepared using 7 and 14 equivalents of metal nitrate per mol of G3 with the aim of establish the maximum amount of metal cation able to be incorporated. Following the same procedure, the volumes of a 0.1 M $M(NO_3)_n$ solution added were 77 and 154 μL.

## 2.3. Drug loading

**MSN-L.** 50 mg of MSN material were soaked in 10 mL of a LEVO solution in $CH_2Cl_2$ (5 mg/mL) and the suspension was stirred at room temperature (RT) for 16 h in dark conditions. Then, the sample was filtered, gently washed with $CH_2Cl_2$ and dried under vacuum. Finally, for comparative purposes, the material was submitted to an aqueous treatment similar to that described for $M^{n+}$ complexation but in this case using only water in the absence of cations, affording MSN-L sample.

**MSN-G3-L.** Taking into account the size of the G3, the antibiotic loading was performed before the external functionalization of the MSN material with G3-Si(OEt)$_3$, which it is performed in the same loading media. For that, 250 mg of vacuum dried MSN (for each sample) were suspended in 25 mL of a LEVO solution in dry $CH_2Cl_2$ (10 mg/mL) and stirred at RT for 16 h in darkness. To these suspensions a solution of the freshly prepared G3-Si(OEt)$_3$ (amounts and conditions as described above for the functionalization of 250 mg of MSN with G3) was added and the stirring was maintained for another 16 h at RT in the absence of light. Then, the samples were centrifuged, gently washed with $CH_2Cl_2$ and dried under vacuum. Finally, for comparison purposes, as above commented, sample was submitted to a water treatment, affording MSN-G3-L.

**MSN-G3-L-$M^{n+}$.** The same procedure detailed for MSN-G3-$M^{n+}$ was carried out to generate MSN-G3-L-$M^{n+}$ starting from the antibiotic loaded MSN-G3-L materials.

## 2.4. *In vial* release assays

Cation release from MSN-G3-$M^{n+}$ nanosystems was assessed in water. For that, 5 mg of MSN-G3-$M^{n+}$ material was suspended in 3 mL of $H_2O$ and kept in an orbital shaker at 37 °C for 1 h. Then, the nanoparticles were centrifuged and the medium renewed with 3 mL of $H_2O$. The isolated supernatant was filtered through a syringe filter (0.2 μm) and the procedure repeated. The supernatants collected at times 1, 3, 6, 24 and 240 h were analysed by ICP-AES. The maximum possible concentration value obtained if the total amount of metal cation were released in the experiment would be 7.8 mg/L for $Zn^{2+}$ and 6.7 mg/L for $Ag^+$ (data based on EDS analysis of materials, Table S1).

The kinetic studies of antibiotic release were carried out in PBS at 37 °C and physiological pH 7.4. A double-chamber cuvette with two different compartments (sample and analysis) was employed for the experiments. The compartments are separated by a dialysis membrane (12 kDa molecular weight cut-off) that only allows the antibiotic diffusion. Briefly, 0.5 mL of a suspension of the loaded MSN materials in PBS (2 mg/mL) was placed in the sample compartment and 1.5 mL of fresh PBS were placed in the analysis compartment. The volume of PBS located in the analysis compartment was renewed at each measurement time. The amount of antibiotic released was measured by fluorescence spectroscopy using a Biotek Powerwave XS spectrofluorimeter (version 1.00.14 of the Gen5 program). For the LEVO analysis, $\lambda_{ex}$ = 292 nm and $\lambda_{em}$ = 494 nm were used in the spectrofluorimeter, and the calibration curve was established in a concentration range of 0.003 to 10 mg/mL.

## 2.5. Microbiological assays

**Bacteria culture.** *Escherichia coli* (*E. coli* ATCC 25922 laboratory strain) as Gram-negative bacteria model was used for the assays. In this case, the LEVO-*E.coli* antibiotic-bacteria binomial has been used as a model due to its specificity. Nevertheless, antibiotics of different families can be loaded into MSNs depending on the target bacteria [28]. *E. coli* bacteria culture was carried out by inoculation in Luria-Bertani broth (LB) and incubated for 3 h at 37 ºC with orbital shaking at 200 rpm. After culture, bacteria were centrifuged for 10 min at 3,500 g at 22 °C. The supernatant was then discarded, and the pellet was washed three times with sterile PBS. The bacteria were then suspended and diluted in PBS to obtain a concentration of $2\times10^9$ bacteria/mL and then a dilution with the corresponding broth is performed to get $2\times10^6$ bacteria/mL. Bacteria concentration was determined by spectrophotometry using a visible spectrophotometer (Photoanalizer D-105, Dinko instruments). Antimicrobial effect tests were carried out both in planktonic state and in preformed biofilms by counting the colony forming units (CFUs) in agar after the exposure of bacteria to the nanosystems. A stock nanoparticle suspension (500 µg/mL) was prepared in PBS 1x by combining vortex and ultrasound and dilutions to the desired concentration were prepared with the corresponding broth taking care to homogenize the suspension by shaking at the time of taking each aliquot.

**Biofilm growth.** *E. coli* biofilms were previous developed onto round cover glasses, impregnated with poly-L-lysine, by placing them into 24-well plates (P-24, CULTEK) and adding 1 mL of a bacteria suspension of $10^6$ bacteria per mL in LB with 0.2% sucrose to favour the robust formation of the biofilm. The plate is maintained 48 h at 37 ºC and orbital stirring at 100 rpm, adding 0.5 mL of fresh medium after the first 24 h. After 48 h, each well was gently washed twice with 1 mL of PBS 1× buffer solution under aseptic conditions to eliminate medium and unbound bacteria.

**Antimicrobial effect of MSN-G3-M$^{n+}$ in planktonic bacteria.** To determine the antimicrobial effect of the different MSN-G3-M$^{n+}$, 0.5 mL of bacteria suspension at $2\times10^6$ bacteria/mL in broth and 0.5 mL of nanoparticle suspension in broth at double the concentration assayed were added to a well in 24-well plates. MSN-G3-Ag$^+$ was assayed at concentrations 1, 2.5, 5 and 10 µg/mL and MSN-G3-Zn$^{2+}$ at 10, 30, 60 and 90 µg/mL. The plates were then incubated at 37 °C with orbital shaking at 200 rpm for 16 h. The presence or not of bacteria, as well as their quantification, was determined after incubation by counting the CFUs using the drop plate method in agar plates [36]. Two serial dilutions in PBS 1x (1:100 and 1:1000) of the bacteria exposed to the nanosystems were made and five drops (10 µL x 5 times) of each solution were inoculated in Tryptic Soy Agar (TSA) plates divided into 3 sectors, which were incubated for 16 h at 37 ºC. The mean count of the 5 drops of each dilution and the average counting for all dilutions was calculated following the procedure described in reference [37]. CFUs/mL compared with bacteria without treatment after 16 h as control was determined. MSN and MSN-G3 nanosystems were used as controls for materials. Data are mean ± SD of three independent experiments.

**Antimicrobial effect of MSN-G3-L-M$^{n+}$ nanosystems against *E. coli* biofilms.** The antimicrobial activity of the different nanosystems on the biofilm was evaluated through two experiments: on one hand, a quantitative assay was carried out calculating the reduction of CFU/mL to assess biofilm viability and, on the other hand, a qualitative analysis of the matrix, living cells and dead cells of each biofilm was performed by confocal laser scanning microscopy (CLSM).

Biofilm viability assay. *E. coli* biofilms were obtained after an incubation period of 48 h as described above and, then, 1 mL of a suspension of the nanosystems in LB in the desired concentration was added.

The exposure of preformed biofilms of *E. coli* to the nanosystems was performed overnight at 37 ºC under orbital stirring at 100 rpm. After incubation, the medium was removed from the wells, which were washed once with 1 mL of sterile PBS 1× and another mL of fresh PBS 1× was added. Subsequently, mechanical disruption of the biofilm was performed with a pipette tip for 30 s and sonication was applied for 10 minutes in a low-power bath sonicator to break and disperse the biofilm in a total volume of 1 mL of PBS 1×. The presence or not of bacteria, as well as their quantification, was determined by counting the CFUs using the drop plate method in LB-agar plates as described above. The dilutions used were 1:100, 1:1000 and 1:10,000 in PBS 1x. All tests were performed in triplicate with their respective controls.

Confocal laser scanning microscopy assay: *E.coli* biofilms were treated with 5 µg/mL or 60 µg/mL of different nanosystems in PBS during 90 min of incubation at 37 ºC. Then, they were washed three times with sterile PBS 1× and 0.5 mL of LB medium was added. Then, 3 µL (1:1 propidium iodide/SYTO) of the Live/Dead Bacterial Viability Kit were added and 5 min later, 5 µL/ml of a calcofluor solution were also added to stain the mucopolysaccharides of the biofilm (extracellular matrix) in blue. Both reagents were incubated for 15 min at RT. Controls containing bacteria biofilm without treatment were also stained. Biofilms were examined in an Olympus FV1200 confocal microscope and six photographs (60× magnification) were taken of each sample. Confocal images were evaluated and quantified using the ImageJ Fiji software (National Institute of Health, Bethesda, MD). All images are representative of three independent experiments.

**2.6. *In vitro* cell assays**

**Cell culture.** Murine pre-osteoblastic cell line (MC3T3-E1) was maintained in alpha modified Eagle's medium (α-MEM). The culture media was supplemented with 2 mM glutamine, 10% fetal bovine serum (FBS) and 1% penicillin/streptomycin at 37 ºC under atmosphere conditions of 95% humidity and 5% $CO_2$. For performing the cell assays, cells were seeded and incubated for 24 h allowing for cell attachment before exposure to the nanosystems.

**Cell uptake assay.** MC3T3-E1 cells were seeded in 6-well plates ($1 \times 10^6$ cell/well) and cultured in the presence of the different zinc-containing MSNs nanosystems for 2 h at 37 °C and 5% $CO_2$. For control experiments, cells were incubated without nanoparticles. After this time, cells were trypsinized and centrifuged at 100 g for 5 min. The cell pellet was resuspended in 150 µL of PBS and 150 µL of Trypan Blue. Trypan blue (TB 0.4%) was added at that time to quench the fluorescence of the MSNs adhered in the outside membrane of the cell, as TB cannot penetrate the membranes of living cells. The mixture was incubated for 5 min at RT protected from light. Analysis of the fluorescein-marked nanoparticles internalized by the living cells was performed on a FACSCan (Becton Dickinson) flow cytometer. The percentage of cells that had internalised nanoparticles was quantified as the fraction of fluorescein positive cells among the counted number of live cells.

**Cytotoxicity assay.** For the cell viability assay, MC3T3-E1 cells were seeded on 96-well plates 24 h prior to the experiment (5000 or 10000 cells/well). After attachment, cells were exposed to the different nanosystems for 2 h at 37 °C and 5% $CO_2$. Subsequently, cells were washed with PBS several times and cultured in fresh medium for 24 h or 96 h. After this time, cell viability was measured by using Thiazolyl

Blue Tetrazolium Bromide following the manufacturer's instructions. The method is based on the fact that only living cells can reduce XTT tetrazolium by an active mitochondrial dehydrogenase enzyme, producing blue crystals of insoluble formazan that can be quantified colorimetrically. For that, 20 µL of 3-(4,5-dimethyl-thiazol-2-yl)2,5-diphenyl tetrazolium bromide (MTT, 5 mg/mL) were added to each well after the selected time and incubated for 4 h at 37 ºC. Then, the MTT solution was removed and 100 µL of dimethyl sulfoxide were added to dissolve the insoluble purple formazan crystals. Finally, the absorbance at 570 nm was measured using a microplate reader (Sinergy 4, BioTek, USA). Cell viability is expressed using untreated cells as control. Data are means ± SD of three independent experiments. To determine the significance, Student's tests were performed. $P < 0.05$ was considered significant. The nanosystems were assayed taking into account the cation and the absence or the loading of LEVO, organized in the following groups and concentrations: *i)* MSN (10 µg/mL), MSN-G3 (10 µg/mL) and MSN-G3-Ag$^+$ (1, 2.5, 5, 10 µg/mL); *ii)* MSN (10 µg/mL), MSN-L (10 µg/mL), MSN-G3-L (10 µg/mL) and MSN-G3-L-Ag$^+$ (1, 2.5, 5, 10 µg/mL); *iii)* MSN (90 µg/mL), MSN-G3 (90 µg/mL) and MSN-G3-Zn$^{2+}$ (10, 30, 60, 90 µg/mL); *iv)* MSN (90 µg/mL), MSN-L, MSN-G3-L (90 µg/mL) and MSN-G3-L-Zn$^{2+}$ (10, 30, 60, 90 µg/mL).

**Osteoblastic differentiation assay.** Pre-osteoblastic MC3T3-E1 cells were seeded in 6-well plates (10,000 cell/well) in α-MEM medium containing 10% FBS and 1% penicillin/streptomycin. Osteoblastic differentiation was performed with α-MEM medium supplemented with 10 mM β-glycerophosphate and 50 µg/mL ascorbic acid as osteogenic medium. Cells were stimulated with the differentiation medium (DM control) in the absence or presence of the zinc-containing MSNs at different concentrations for 7 and 14 days.

**Mineralization.** Patterns of matrix mineralization in MC3T3-E1 cells were analysed using Alizarin Red S staining. MC3T3-E1 cells were cultured with α-MEM or osteogenic medium (DM control) in the absence or presence of the nanosystems for 7 and 14 days. Then, cells were washed with PBS three times and fixed with 70% ethanol for 1 h and stained with Alizarin Red S for 30 min. After this time, cells were washed with distilled H$_2$O to remove nonspecific staining. The calcium compound deposited on the wells was visually scored by a scanner (BIOTEK). Finally, to quantify the matrix mineralization, the stained samples were eluted with cetylpyridinium chloride (10% w/v) in PBS (pH = 7) and the absorbance was measured at 620 nm.

**Alkaline phosphatase (ALP) activity assay.** The alkaline phosphatase (ALP) activity was assayed with a reagent kit (1001130, Spinreact). MC3T3-E1 cells were cultured with α-MEM or osteogenic medium in the presence or absence of the nanosystems for 14 days. On days 7 and 14, the ALP secreted into the culture medium was measured with this kit following the manufacturer's instructions. Briefly, the medium was removed from the cells and 5 freezing cycles were performed at −80 °C for 30 min. After this time, 1 mL of the kit mixture was added to the cells, incubated for 30 min at 100 rpm and the reaction was stopped with 200 µL of NaOH (2 M). Finally, the absorbance was measured at 405 nm.

**2.7. Statistical analysis**

*In vitro* data are expressed as mean ± standard deviations of three independent experiments. Unpaired two-tailed *Student's t-test* was used to determine the statistical significance. In all of the statistical evaluations, $P < 0.05$ was considered as statistically significant. Statistical analyses for antibiotic release, microbiological results and cell assays were performed using the Graphpad Prism program (Graphpad software, USA).

## 3. RESULTS AND DISCUSSION

### 3.1. Materials synthesis and characterization

The nanosystems here presented consist in a dendrimer functionalized MSN nanoplatform containing both an antimicrobial cation ($M^{n+}$) onto the external surface and an antibiotic, namely LEVO, inside of the mesoporous structure. For their synthesis, polypropyleneimine dendrimers of third generation (G3) were covalently bonded to the external surface of MSNs (MSNs-G3) in a first step (Fig. S2) [30]. Subsequently, $M^{n+}$ cations were tethered on the dendritic branches through their complexation with the tertiary amine groups of the dendrimer to afford MSNs-G3-$M^{n+}$ materials (Fig. 1A). The complexation of the cationic metals ($M^{n+}$ = $Ag^+$ or $Zn^{2+}$) was performed in water using the metal nitrates as cation sources. To incorporate the maximum amount of metal cation in the material, a study was firstly carried out taking into account that the dendrimer in the MSN-G3 material possesses 14 moles of tertiary amines per mole of G3 dendrimer. Primary amines were not taken into account due to their protonation equilibrium in water media, which makes them unable to coordinate metal cations (Eq. 1 and 2) [38,39]. The protonation of G3 was confirmed by the appearance of the $\nu_{NH}$ vibration band of $NH_3^+$ at 2926 cm$^{-1}$ in FTIR spectrum of samples after cations incorporation (Fig. S3). Different stoichiometries of metal nitrate per mole of G3 (3.5, 7 and 14 equivalents) were used. The analysis of the M/Si molar ratio obtained by EDS analysis showed that above the 3.5 stoichiometry it is not observed a greater amount of metal cation incorporated into the material (Table S1). In view of these results, we chose the 3.5 stoichiometry to prepare the MSN-G3-$M^{n+}$ materials. Cation incorporation was performed in the last step of synthesis for nanosystems either antibiotic unloaded or loaded (Fig. 1). The antibiotic was loaded into MSN following an impregnation method [28,40] simultaneously to the grafting of G3 alkoxysilane on the external surface so the accessibility of LEVO to the inner pores is not hindered (Fig. 1B). For this, the dendrimer anchorage to the external surface of MSNs was carried out in the loading media, thus ensuring the presence of antibiotic molecules in the mesopores of MSN to rule out that steric hindrance prevents or hinders the diffusion of antibiotic molecules into the mesopores. Then, the complexation of metal cations was accomplished by soaking the resulting material into an aqueous $M(NO_3)_n$ solution, where $M^{n+}$ stands for $Ag^+$ or $Zn^{2+}$, affording MSNG3-L-$Ag^+$ and MSN-G3-L-$Zn^{2+}$ materials, respectively.

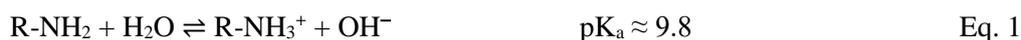

R-NH$_2$ + H$_2$O ⇌ R-NH$_3^+$ + OH$^-$     pK$_a$ ≈ 9.8     Eq. 1

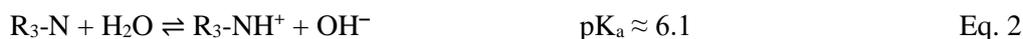

R$_3$-N + H$_2$O ⇌ R$_3$-NH$^+$ + OH$^-$     pK$_a$ ≈ 6.1     Eq. 2

Fig. 2 shows TEM images corresponding to pristine MSN nanoparticles and the dendrimer functionalized counterparts after cation complexation, MSN-G3-$Ag^+$ and MSN-G3-$Zn^{2+}$, respectively. All TEM images reveals nanoparticles with uniform quasi-spherical morphologies and an average particle size of *ca.* 180 nm, exhibiting a mesoporous honeycomb arrangement. Furthermore, XRD analysis (Fig. S4A) shows diffraction patterns corresponding to the typical MCM 41 2D hexagonal structure (*p6mm* plane group) [41]. This reveal that the functionalization and subsequent cation complexation do not alter neither the morphology nor the mesoporous order of the nanoparticles. Main difference between TEM images of bare MSNs and the MSN-G3-$Ag^+$ and MSN-G3-$Zn^{2+}$ nanosystems is the observed darker area in the external surface part of cation bearing ones. This fact could be attributed to the higher intensity of an electron rich region due to the cations complexation in the surface of the nanosystems, *i.e.*, where the polyamine dendrimer is located. The presence of $Ag^+$ or $Zn^{2+}$ was also confirmed by EDS, which provided atomic percentage values of 0.30 ± 0.02 and 0.5 ± 0.1 for silver and zinc in MSN-G3-$Ag^+$ and MSN-G3-$Zn^{2+}$ nanosystems, respectively (see Table S1). The incorporation of metal ions into the structure is availed by the presence of $NO_3^-$ ions in the resulting nanosystems, as confirmed by FTIR throughout the significant increase in the relative intensity of the vibration band at 1382 cm$^{-1}$ (Fig. S3) [42].

ζ-potential measurements in water media confirm the dendrimer incorporation in the MSNs surface because a drastic change from negative to positive values is produced after functionalization. Values change from −22.8 mV in bare MSN to *ca.* +40 mV once the polyamine dendrimer is anchored to the surface. These values are due to deprotonation of silanol groups in MSN surface (Eq. 3) and protonation of primary amine groups in G3 dendrimer after functionalization (Eq. 1) [38,39,43]. The value for MSN-G3 is so high positive that it is preserved after the complexation of the cations without being increased (see Table 1 and Fig.S5).

$$R\text{-}Si\text{-}OH + H_2O \rightleftharpoons R\text{-}SiO^- + H_3O^+ \qquad pK_a \approx 6.8 \qquad Eq.\ 3$$

The hydrodynamic diameter of the nanosystems was measured by DLS in water medium, displaying monomodal number-based hydrodynamic size distributions for all samples (Table 1 and Fig. S5). Functionalization with the G3 dendrimer and subsequent incorporation of the cations barely influences the hydrodynamic size of the initial MSNs because all the materials exhibit ζ-potential values in the zone of colloidal stability [44] and therefore electrostatic repulsion is guaranteed. Furthermore, the hydrodynamic size of MSN-G3 material is hardly affected by the concentration of nanoparticles in the suspension, probably due to the additional effect of steric repulsion due to the incorporation of macromolecules in the MSNs surface (see table in Figure S5).

## 3.2. *In vial* release assays

To assess the stability of the nanosystem regarding the possible lixiviation of metal ions in the biological media, a cation release test was performed. After soaking the MSN-G3-M$^{n+}$ materials in water media the supernatant was analysed by ICP-AES at different times (Table S2). Although the zinc cation is slightly released during the first 6 h of the assay, silver is not released at any time from the nanosystem. Furthermore, release of Zn$^{2+}$ decreases with the immersion time, until no more quantity is released thereafter 6 h. The resulting cumulative value for the Zn$^{2+}$ ion is 0.588 mg/L, which represents only a 7.5% of the maximum amount of Zn$^{2+}$ possible to be released from the sample (7.8 mg/L). These results show that cations would remain attached to the surface of the nanosystems during the biological assays, pointing out to a mechanism of action involving the whole nanosystem.

To evaluate the release behavior of the different nanosystems, namely MSN-G3-L, MSN-G3-L-Zn$^{2+}$ and MSN-G3-L-Ag$^+$, under physiological conditions (PBS 1x, pH = 7.4), antibiotic release experiments were carried out "*in vial*." Fig. 3 shows the LEVO release profiles for the different materials after 95 h of testing expressed as micrograms of LEVO released per mg of material. All release profiles are characteristic of the typical diffusion pattern of MSN materials. Thus, the release profiles were fitted to the first order kinetics (Eq. 4) [45]:

*Y = A (1 − e$^{-kt}$)*      *Eq. 4*

being *Y* the amount of LEVO released (µg LEVO/mg material) at *t* time (h), *A* the maximum amount of LEVO released (µg LEVO/mg sample) and *k* the release rate constant. The parameters resulting from kinetics fitting are shown in Table 2. As it is observed, there are significant differences in the *A* values for the different samples, being 23.3 49.1 and 220.8 µg/mg, for MSN-G3-L, MSN-G3-L-Ag$^+$ and MSN-G3-L-Zn$^{2+}$, respectively. Such dissimilarities can be attributed to the different extent of LEVO leakage taking place during the final soaking step in aqueous $M(NO_3)_n$ ($M^{n+}$ = $Ag^+$, $Zn^{2+}$) solutions or cation-free water, following LEVO impregnation and G3 grafting. During such process, the protonated primary amino groups in the external surface of G3 would promote the interaction with LEVO molecules, provoking partial drug departure out the mesopores [30,46]. In the absence of cations thanks to its open-flexible structure, the G3 would easily rearrange to procure interaction of its protonated primary amino end-groups with LEVO molecules, prompting drug diffusion from the inner mesopores to water medium [30]. On the contrary, the complexation of $M^{n+}$ with the tertiary amines in the inner core of the G3 would decrease the flexibility of the dendrimer [47,48], diminishing the possibilities of interaction of LEVO with the dendritic branches and creating additional barriers for the diffusion of the drug out the mesopores. This effect is more evident in the case of MSN-G3-L-$Zn^{2+}$, not only ascribed to the higher metal cation content (see Table S1) but also owing to the trend of $Zn^{2+}$ to form complexes with higher coordination index than $Ag^+$ does. Indeed, it is well-known the trend of zinc cations to form stable chelates showing tetrahedral configuration with polydentate ligands [49]. This would increase rigidity of the resulting chelate ring, slowing down the premature cargo departure and leading to higher LEVO content in MSN-G3-L-$Zn^{2+}$ than in MSN-G3-L. However, in the case of MSN-G3-L-$Ag^+$, the trend of silver cation to form lineal complexes would produce a relatively more flexible core complex that would allow LEVO leakage in a similar fashion to cation-free MSN-G3-L sample [50]. This finding is in good agreement with the noticeable differences observed in the release rate ($k$) values, being 1.00, 0.66 and 0.51 $h^{-1}$ for MSN-G3-L, MSN-G3-L-$Ag^+$ and MSN-G3-L-$Zn^{2+}$, respectively. The slower antibiotic release observed for MSN-G3-L-$Zn^{2+}$ sample would agree with the decrease in the flexibility of the G3 dendrimer after chelation process, as above discussed, whose rearrangement to procure interaction with LEVO molecules would be more restricted, even though, eventually, not impeding the antibiotic release.

### 3.2. Microbiological assays

The bactericidal effect of the different nanosystems was evaluated in *E. coli* biofilms, selected as Gram-negative bacteria model and considering its sensitivity to the chosen antibiotic, LEVO [28]. Firstly, the antimicrobial efficiency and most effective doses of the cation containing nanosystems (MSN-G3-$M^{n+}$) were assessed in planktonic *E. coli* bacteria (see the discussion of the results in the Supporting Information and Fig. S6-S8). The results derived from these studies in planktonic state show a high antimicrobial efficacy for nanosystems incorporating the cations at concentrations of 2.5 µg/mL for MSN-G3-$Ag^+$ and 30 µg/mL for MSN-G3-$Zn^{2+}$. Then, the MSN-G3-L-$M^{n+}$ nanosystems were assayed on preformed biofilms to evaluate the combined effect due to the presence of both the cation and the antibiotic in the same nanoplatform.

Preliminary confocal microscope studies were done to evaluate the direct effect of MSN-G3-$M^{n+}$ nanosystems onto preformed biofilm (Fig. 4) The control biofilm shows a typical structure composed by a mat of live bacteria (green) covered with a protective polysaccharide layer (blue). However, for biofilms incubated for 2 h with MSN-G3-$Ag^+$ (5 µg/mL) or MSN-G3-$Zn^{2+}$ (60 µg/mL) samples, a remarkable reduction of the biofilm area was observed, together with the presence of dark gaps and small bacterial colonies with isolated bacteria along the entire surface. These results indicate that these nanosystems cause disruption of biofilm, as it has been previously reported for cationic amine functionalized MSN materials [30,46]. This disruption effect is more pronounced in the silver-containing nanosystems, being observed smaller colonies *ca.* 20 µm with almost no protective coating. The differences in antimicrobial efficacy between both MSN-G3-$M^{n+}$ nanosystems can be attributed to the greater antimicrobial effect of $Ag^+$ *versus* $Zn^{2+}$ [16] and are consistent with the antimicrobial effect in the planktonic state (see Supporting Information).

Once tested the effect of cation incorporating samples (MSN-G3-$M^{n+}$) on biofilm, the bactericidal effect onto *E. coli* biofilms of nanosystems loaded with both the antibiotic LEVO and the cation (MSN-G3-L-$M^{n+}$) was evaluated to define the combined effect of cation and antibiotic in a single nanosystem.

For this purpose, preformed *E. coli* biofilms were exposed to different concentrations of nanosystems, 2.5 and 5 µg/mL for silver containing samples and 30 and 60 µg/mL for zinc containing samples, based on the bactericidal efficacy found for these doses in the planktonic state. Nanosystems containing only one of the microbicide agents, *i.e.* the cation (MSN-G3-$M^{n+}$) or the antibiotic (MSN-G3-L), produce a notable reduction of biofilm. However, values above 99% reduction are reached with nanosystems incorporating both agents simultaneously (Fig. 5, and S9-S11). Hence, these results show the combined action of both components (metal cation and antibiotic) in a single nanosystem. In the case of silver nanosystems, the biofilm inhibition percent for MSN-G3-L and MSN-G3-$Ag^+$ at 5 µg/mL was 97.4 and 90.8%, respectively, achieving a total of 99.6% for the multicomponent nanosystem MSN-G3-L-$Ag^+$. The multicomponent effect was more pronounced for the zinc nanosystem, where the biofilm reduction for MSN-G3-L and MSN-G3-$Zn^{2+}$ at the assayed doses were 97.4 and 90.8%, reaching a considerable reduction of 99.9 % for MSN-G3-L-$Zn^{2+}$ at 60 µg/mL.

### 3.4. *In vitro* cell studies in preosteoblastic cell line

**Cell uptake**

Cell uptake assays were carried out to check whether the nanosystems internalize into cells. Figure 6 shows the percentage of MC3T3-E1 cells that have internalized the MSN-G3-$Zn^{2+}$ and MSN-G3-L-$Zn^{2+}$ nanosystems at different doses (10, 30 and 60 μg/mL) after 2 h of exposure (see also Fig. S12). MSNs and MSN-L at 60 μg/mL were internalized in approximately 70 and 61% of the cells, respectively. When the dendrimer is attached the cell uptake increases *ca.* 25%, being 95 and 85% for MSN-G3 and MSN-G3-L, respectively. This fact is due to the change in ζ-potential from negative to positive values after polycationic dendrimer functionalization of MSNs [30,51]. Therefore, the uptake of positively charged nanoparticles is more favoured that negatively charged bare silica nanoparticles considering that the resting potential of the cell membranes is usually negative. The decrease in the internalization degree for LEVO-containing samples, which exhibit a less positive surface charge value than antibiotic-free-nanosystems (*vide supra*), supports this fact. On the other hand, when the $Zn^{2+}$ cation is incorporated in the nanosystem ζ-potential values are not altered (see Table 1), which explains very similar cell uptake values of *ca.* 90% for MSN-G3 and MSN-G3-$Zn^{2+}$ materials at 60 μg/mL. The cell uptake of the MSN-G3-$Zn^{2+}$ and MSN-G3-L-$Zn^{2+}$ material is dose dependent, reaching maximum values of 94 and 87%, respectively. These findings, *i.e.* the lack of significant cation release together with improved cell uptake of nanosystems, support that the mechanism of action of cations incorporated in the MSN-G3 nanoplatform is different from free cations in solution because $Zn^{2+}$ cations incorporated in the MSN-G3 nanoplatform are effectively transported inside cells through the internalization of the nanosystem.

**Cell viability**

Cell viability was evaluated in the preosteoblastic cell line MC3T3-E1 exposed to MSN-G3-L-$M^{n+}$ nanosystems for 24 and 96 h (Fig. 7 and Fig S13). Different concentrations in the range where each metal-containing nanosystem exhibits bactericide effect were tested. Doses in the 10-90 μg/mL range were assayed for the zinc-containing nanosystems. Non-cytotoxic effect is observed for the MSN-G3-L-$Zn^{2+}$ nanosystem up to 60 μg/mL, being the cell viability values 77-93% for 24 h and 79-87% after 96 h, respectively. Only at the highest tested dose of 90 μg/mL the nanosystem showed toxic effect, with viability values of 25% and 49% for 24 h and 96 h, respectively. The obtained results indicate that doses exhibiting bactericide effect were cytocompatible, and accordingly, we selected doses up to 60 µg/mL to evaluate the influence of the zinc-containing nanosystems in the osteogenic effect. It is worth to mention that the free cation (last bar in Fig. 7) shows no cytotoxicity as observed for the maximum doses tested at both 24 and 96 h, being the lowest value for cell viability *ca.* 79% for $Zn^{2+}$ at 24 h. Nevertheless, the higher cytotoxicity observed for the $Zn^{2+}$-containing nanosystem at the dose of 90 µg/mL could be ascribed to the fact that cations incorporated into the nanosystem through complexation with the dendrimer are not released in the aqueous medium (*vide supra*), together with the internalization of the nanosystems into cells, as confirmed with the cell uptake assays. Moreover, it is important to remark that the cell viability of LEVO-loaded MSN-G3-L-$Zn^{2+}$ nanosystem improves with respect to that of LEVO-free MSN-G3-$Zn^{2+}$. Hence, MSN-G3-$Zn^{2+}$ initially shows viability values below 60%, although it raises up acceptable cell viability values after 96 h of incubation. The better results of MSN-G3-L-$Zn^{2+}$ can be explained due to a decrease in the internalization degree compared to the LEVO-free nanosystem (*vide supra*). The lower internalization degree of LEVO-loaded nanosystems is ascribed to a decrease in the surface charge due to the antibiotic incorporation. Thus, the ζ-potential values were 41.2 mV and 36.1 mV for MSN-G3-$Zn^{2+}$ and MSN-G3-L-$Zn^{2+}$, respectively.

Similar cell viability behavior was found for silver-containing nanosystems in the tested concentration range of 1-10 µg/mL (Fig. S13). MSN-G3-$Ag^+$ produces a decrease in cell viability at 24 h below the 60% in a dose dependent manner, reaching a decrease of up to 37% when using the concentration of 10 µg/mL. However, the LEVO-loaded MSN-G3-L-$Ag^+$ nanosystem maintains cell viability above 78% for all the concentrations. It should be noted that cell viability values at 96 h show higher values than those found at 24 h, with no dose being cytotoxic (all above 82%) for the nanosystem that contains both the cation and the antibiotic. Notably, the MSN-G3-L-$Ag^+$ nanosystem presents values of 93% for the dose 1 µg/mL which have a bactericidal effect for *E. coli* in planktonic state, as well as 88 and 86% for doses effective in biofilm, 2.5 and 5 µg/mL, respectively.

**Osteogenic effect assays**

The effect of zinc-containing nanosystems on the osteogenic differentiation was evaluated in the preosteoblast cell line MC3T3-E1 through the measurement of bone cell mineralization and alkaline phosphatase activity. Patterns of matrix mineralization were examined at days 7 and 14 by staining calcium compounds deposited on the wells with Alizarin Red S which were visually scored by a scanner (Fig. 8A) and quantified through the absorbance at 620 nm (Fig. 8B). The osteogenic differentiation was evident after alizarin red staining for all samples except the control culture, because the redness of the nodules indicates the presence of mineralized compartments as a result of the osteogenic treatment. The quantification at day 7 shows that the zinc containing nanosystem MSN-G3-L-$Zn^{2+}$ generates a higher amount of mineralized nodules than the obtained with the non-containing $Zn^{2+}$ controls. Furthermore, the behaviour is dose dependent as the amount of calcium deposits rises up when the nanosystem concentration increases from 10 to 60 µg/mL.

As expected, at day 14 the amount of mineralization nodules is greater than at day 7 for each material, and the MSN-G3-L-$Zn^{2+}$ even doubles up the calcium deposits with the intermediate dose of 30 µg/mL and multiply them by three with the highest dose of 60 µg/mL. Therefore, when the formation of mineral nodules is evaluated at longer times, the differences between the controls without $Zn^{2+}$ and the $Zn^{2+}$ containing nanosystems are more pronounced. Interestingly, when comparing the amount of mineral nodules for the MSN-G3-L-$Zn^{2+}$ nanosystem at 60 µg/mL it is twice (day 7) or trice (day 14) the obtained value with the free $Zn^{2+}$ cation in the same amount contained in the nanosystem at such dose.

This result maybe explained with the fact that $Zn^{2+}$ is fairly released from the nanosystem and therefore it is transported to the cell interior through the cell uptake of nanosystem. Therefore, cations incorporated in the MSN-G3 nanoplatform are effectively transported inside cells through cell uptake of the nanosystem (see below) unlike free cations in solution.

Alkaline phosphatase (ALP) is an enzyme that participates in the process of bone formation and mineralization at the early stage of osteogenic differentiation [52]. Therefore, the activity of ALP enzyme was measured both in the culture medium at days 7 and 14 (Fig. 8C) and within cells at day 14 (Fig. 8D). Extracellularly, ALP activity is similar at day 7 and 14 for each one of the controls, respectively. There is a dose and time-dependent increase in the ALP activity for the MSN-G3-L-$Zn^{2+}$ nanosystem compared to non-containing $Zn^{2+}$ materials as well as for the free cation. The highest ALP activity was detected for MSN-G3-L-$Zn^{2+}$ at 60 μg/mL.

Therefore, matrix mineralization and ALP activity results indicate the stimulatory effect of endocytosed MSN-G3-L-$Zn^{2+}$ nanosystem on the differentiation of MC3T3-E1. The 60 μg/mL dose generates osteogenic differentiation without compromising viability as well as it is a dose that effectively reduces bacterial biofilm.

The antibiotic unloaded nanosystems were also assayed in the same terms and conditions (see Fig. S14) resulting in the same above discussed behaviour. This result indicates that the presence of LEVO would not decrease the osteogenic effect of the zinc-containing nanosystem when used as bactericide in a bone infection scenario where also bone regeneration is needed.

## 4. CONCLUSIONS

MSN-based nanosystems with dual antimicrobial and osteogenic capability have been designed by a simple and versatile synthesis approach. These MSN nanosystems, containing two antimicrobial agents, LEVO and $Zn^{2+}$, have been synthetized by external functionalization of MSNs with a polycationic dendrimer (MSNs-G3), which favours its internalization inside the bacteria and lead the complexation with metal ions through the amines of the dendrimer. The nanosystems afford a notable enhancement of the antibiofilm effect (above 99%) than both components separately as well as osteogenic capability with positive effect on the osteoblastic differentiation and preserved cell viability. One of the strengths of this work is the simplicity and versatility of the methodology used, allowing the incorporation of a wide range of biologically multiactive metal ions by simple soaking into the metal cation solution. The selection of the cation may be done depending on clinical needs. Thus, the treatment of different pathogenic bacterial biofilms can be tailored by simply changing the loaded antibiotic attending to the clinical needs. The results derived from the current research opens the gates towards the development of personalized therapies for the management of bone infection. Hence, research exploring the intra-osseous injection of a nanoparticle suspension at the bone infection site as local administration route, as well as the *in vivo* therapeutic effect and biodistribution, is ongoing and will be the object of further investigations.

**Conflicts of interest**

Authors declare no conflicts of interest.

**ACKNOWLEDGEMENTS**

This study was supported by the European Research Council ERC-2015-AdG (VERDI) Proposal No. 694160 and the Ministerio de Ciencia e Innovación MAT2016-75611-R and PID2020-117091RB-I00 grants.

**REFERENCES**

[1] C.R. Arciola, D. Campoccia, L. Montanaro, Implant infections: adhesion, biofilm formation and immune evasion, Nat. Rev. Microbiol. 16 (2018) 397-409.

[2] D. Davies, Understanding biofilm resistance to antibacterial agents, Nat. Rev. Drug Discov. 2 (2003) 114-122.

[3] F. Götz, *Staphylococcus* and biofilms, Mol. Microbiol. 43 (2002) 1367-1378.

[4] S. Hathroubi, M.A. Mekni, P. Domenico, D. Nguyen, M. Jacques, Biofilms: Microbial Shelters Against Antibiotics, Microb. Drug Resist. 23 (2017) 147-156.

[5] H.-C. Flemming, EPS-Then and Now, Microorganisms 4 (2016) 41.

[6] C. Wagner, G.M. Hänsch, Pathophysiology of implant-associated infections: From biofilm to osteolysis and septic loosening, Orthopade 44 (2015) 967-973.

[7] A.J. Salinas, P. Esbrit, M. Vallet-Regí, A tissue engineering approach based on the use of bioceramics for bone repair.Biomaterials Science 1 (2013) 40-51.

[8] A.J. Salinas, M. Vallet-Regí, I. Izquierdo-Barba, Biomimetic apatite deposition on calcium silicate gel glasses. J. Sol-gel Sci. Tech. 21 (2001), 13-25.

[9] N. Rangavittal, A.R. Landa-Cánovas, J.M. González-Calbet, M. Vallet-Regí, Structural study and stability of hydroxyapatite and β-tricalcium phosphate: Two important bioceramics. J. Biomed. Mater. Res. 51 (2000), 660-668.

[10] J. Pérez-Pariente, F. Balas, J. Roman, A.J. Salinas, M. Vallet-Regí, Influence of composition and surface characteristics on the in vitro bioactivity of $SiO_2-CaO-P_2O_5-MgO$ sol-gel glasses. J. Biomed. Mater. Res. 47 (1999), 170-175.

[11] M. Vallet-Regí, D. Lozano, B. González, I. Izquierdo-Barba, Biomaterials against Bone Infection, Adv. Healthcare Mater. 9 (2020) 2000310.

[12] I. Izquierdo-Barba, M. Colilla, M. Vallet-Regí, Zwitterionic ceramics for biomedical applications, Acta Biomater. 40 (2016) 201-211.

[13] I. Izquierdo-Barba, J.M. García-Martín, R. Álvarez, A. Palmero, J. Esteban, C. Pérez-Jorge, D. Arcos, M. Vallet-Regí, Nanocolumnar coatings with selective behavior towards osteoblast and Staphylococcus aureus proliferation, Acta Biomater. 15 (2015) 20-28.

[14] R. Álvarez, S. Muñoz-Piña, M.U. González, I. Izquierdo-Barba, I. Fernández-Martínez, V. Rico, D. Arcos, A. García-Valenzuela, A. Palmero, M. Vallet-Regí, A.R. González-Elipe, J.M. García-Martín, Antibacterial Nanostructured Ti Coatings by Magnetron Sputtering: From Laboratory Scales to Industrial Reactors, Nanomaterials 9 (2019) 1217.

[15] Q. Ye, W. Chen, H. Huang, Y. Tang, W. Wang, F. Meng, H. Wang, Y. Zheng, Iron and zinc ions, potent weapons against multidrug-resistant bacteria, Appl. Microbiol. Biotechnol. 104 (2020) 5213-5227.

[16] W. Fan, Q. Sun, Y. Li, F.R. Tay, B. Fan, Synergistic mechanism of $Ag^+$-$Zn^{2+}$ in anti-bacterial activity against *Enterococcus faecalis* and its application against dentin infection, J. Nanobiotechnology 16 (2018) 1-16.

[17] A.B. Lansdown, A pharmacological and toxicological profile of silver as an antimicrobial agent in medical devices, Adv. Pharmacol. Sci. 2010 (2010) Article ID 910686 16 pages.

[18] J. O'Neill, Antimicrobial Resistance: Tackling a crisis for the health and wealth of nations, Review on Antimicrobial Resistance 2014.

[19] M. Colilla, M. Vallet-Regí, Targeted Stimuli-Responsive Mesoporous Silica Nanoparticles for Bacterial Infection Treatment, Int. J. Mol. Sci. 21 (2020) 8605.

[20] A.J. Huh, Y.J. Kwon, "Nanoantibiotics": a new paradigm for treating infectious diseases using nanomaterials in the antibiotics resistant era, J. Control. Release 156 (2011) 128–145.

[21] A. Bassegoda, K. Ivanova, E. Ramon, T. Tzanov, Strategies to prevent the occurrence of resistance against antibiotics by using advanced materials, Appl. Microbiol. Biotechnol. 102 (2018) 2075-2089.

[22] R.R. Castillo, D. Lozano, B. González, M. Manzano, I. Izquierdo-Barba, M. Vallet-Regí, Advances in mesoporous silica nanoparticles for targeted stimuli-responsive drug delivery: an update, Expert Opin. Drug Deliv. 16 (2019) 415-439.

[23] M. Vallet-Regí, M. Colilla, I. Izquierdo-Barba, M. Manzano, Mesoporous Silica Nanoparticles for Drug Delivery: Current Insights, Molecules 23 (2018) 47.

[24] M. Vallet-Regí, M. Colilla, I. Izquierdo-Barba, Bioactive mesoporous silicas as controlled delivery systems: application in bone tissue regeneration. J. Biomed. Nanotechnology 4 (2008) 1-15.

[25] M. Gisbert-Garzarán, M. Manzano, M. Vallet-Regí, Mesoporous Silica Nanoparticles for the Treatment of Complex Bone Diseases: Bone Cancer, Bone Infection and Osteoporosis, Pharmaceutics 12 (2020) 83.

[26] M. Vallet-Regí, A. Rámila, R.P. Del Real, J. Pérez-Pariente, A new property of MCM-41: drug delivery system, Chem. Mater. 13 (2001) 308-311.

[27] M. Vallet-Regí, B. González, I. Izquierdo-Barba, Nanomaterials as Promising Alternative in the Infection Treatment, Int. J. Mol. Sci. 20 (2019) 3806.

[28] A. Aguilar-Colomer, M. Colilla, I. Izquierdo-Barba, C. Jiménez-Jiménez, I. Mahillo, J. Esteban, M. Vallet-Regí, Impact of the antibiotic-cargo from MSNs on gram-positive and gram-negative bacterial biofilms, Microporous Mesoporous Mater. 311 (2021) 110681.

[29] M. Martínez-Carmona, I. Izquierdo-Barba, M. Colilla, M. Vallet-Regí, Concanavalin A-targeted mesoporous silica nanoparticles for infection treatment, Acta Biomater. 96 (2019) 547-556.

[30] B. González, M. Colilla, J. Díez, D. Pedraza, M. Guembe, I. Izquierdo-Barba, M. Vallet-Regí, Mesoporous silica nanoparticles decorated with polycationic dendrimers for infection treatment, Acta Biomater. 68 (2018) 261-271.

[31] S. Shruti, A.J. Salinas, G. Lusvardi, G. Malavasi, L. Menabue, M. Vallet-Regí, Mesoporous bioactive scaffolds prepared with cerium-, gallium- and zinc-containing glasses, Acta Biomater. 9 (2013) 4836-4844.

[32] G. Jin, H. Cao, Y. Qiao, F. Meng, H. Zhu, X. Liu, Osteogenic activity and antibacterial effect of zinc ion implanted titanium, Colloids Surf. B 117 (2014) 158–165.

[33] R. Pérez, S. Sanchez-Salcedo, D. Lozano, C. Heras, P. Esbrit, M. Vallet-Regí, A.J. Salinas, Osteogenic effect of ZnO-mesoporous glasses loaded with osteostatin, Nanomaterials 8 (2018) 592.

[34] C. Heras, J. Jiménez-Holguín, A.L. Doadrio, M. Vallet-Regí, S. Sánchez-Salcedo, A.J. Salinas, Multifunctional antibiotic- and zinc-containing mesoporous bioactive glass scaffolds to fight bone infection, Acta Biomater. 114 (2020) 395-406.

[35] J.R. Morones, J.L. Elechiguerra, A. Camachom, K. Holt, J.B. Kouri, J.T. Ramírez, M.J. Yacaman, The bactericidal effect of silver nanoparticles, Nanotechnology 16 (2010) 2346–2353.

[36] B. Herigstad, M. Hamilton, J. Heersink, How to optimize the drop plate method for enumerating bacteria, J. Microbiol. Methods 44 (2001) 121-129.

[37] A.J. Hedges, Estimating the precision of serial dilutions and viable bacterial counts, Int. J. Food Microbiol. 76 (2002) 207-14.

[38] V.A. Kabanov, A.B. Zezin, V.B. Rogacheva, Z.G. Gulyaeva, M.F. Zansochova, J.G.H. Joosten, J. Brackman, Polyelectrolyte behavior of astramol poly (propyleneimine) dendrimers, Macromolecules 31 (1998) 5142-5144.

[39] R.C. van Duijvenbode, M. Borkovec, G.J.M. Koper, Acid-base properties of poly (propylene imine) dendrimers, Polymer 39 (1998) 2657-664.

[40] M. Cicuéndez, I. Izquierdo-Barba, M.T. Portolés, M. Vallet-Regí, Biocompatibility and levofloxacin delivery of mesoporous materials, Eur. J. Pharm. Biopharm. 84 (2013) 115-124.

[41] C.T. Kresge, M.E. Leonowicz, W.J. Roth, J.C. Vartuli, J.S. Beck, Ordered mesoporous molecular sieves synthesized by a liquid-crystal template mechanism, Nature 359 (1992) 710–712.

[42] S.A. Silva, C.Q.F. Leite, F.R. Pavan, N. Masciocchi, A. Cuin, Coordinative versatility of a Schiff base containing thiophene: Synthesis, characterization and biological activity of zinc(II) and silver(I) complexes, Polyhedron, 79 (2014) 170-177.

[43] A. Nieto, M. Colilla, F. Balas, M. Vallet-Regí, Surface electrochemistry of mesoporous silicas as a key factor in the design of tailored delivery devices, Langmuir 26 (2010) 5038-5049.

[44] J.M. Rosenholm, C. Sahlgren, M. Lindén, Towards multifunctional, targeted drug delivery systems using mesoporous silica nanoparticles – opportunities & challenges, Nanoscale 2 (2010) 1870-1883.

[45] F. Balas, M. Manzano, M. Colilla, M. Vallet-Regí, L-Trp adsorption into silica mesoporous materials to promote bone formation, Acta Biomater. 4 (2008) 514-522.

[46] D. Pedraza, J. Díez, I. Izquierdo-Barba, M. Colilla, M. Vallet-Regí. Amine-Functionalized Mesoporous Silica Nanoparticles: A New Nanoantibiotic for Bone Infection Treatment. Biomed. Glasses. 4 (2018) 1-12.

[47] M.S. Diallo, S. Christie, P. Swaminathan, L. Balogh, X. Shi, W. Um, C. Papelis, W.A. Goddard, J.H. Johnson, Dendritic Chelating Agents. 1. Cu(II) Binding to Ethylene Diamine Core Poly(amidoamine) dendrimers in aqueous solutions, Langmuir 20 (2004) 2640-2651.

[48] M.S. Diallo, W. Arasho, J.H. Johnson, W.A. Goddard, Dendritic Chelating Agents. 2. U(VI) Binding to Poly(amidoamine) and Poly(propyleneimine) dendrimers in aqueous solutions, Environ. Sci. Technol, 42 (2018) 1572-1579.

[49] T.C. Amaral, F.B. Miguel, M.R.C. Couri, P.P. Corbi, M.A. Carvalho, D.L. Campos, F.R. Pavan, A. Cuin, Silver(I) and zinc(II) complexes with symmetrical cinnamaldehyde Schiff base derivative: Spectroscopic, powder diffraction characterization, and antimycobacterial studies, Polyhedron 146 (2018) 166-171.

[50] J. Bjerrum, On the Tendency of the Metal Ions toward Complex Formation, Chem. Rev. 46 (1950) 381-401.

[51] B. González, M. Colilla, M. Vallet-Regí, Design of In Vitro Bioactive Hybrid Materials from the First Generation of Amine Dendrimers as Nanobuilding Blocks, Chem. Eur. J. 19 (2013) 4883-4895.

[52] U. Sharma, D. Pal, R. Prasad, Alkaline Phosphatase: An Overview, Indian J. Clin. Biochem. 29 (2014) 269-278.

**TABLES**

**Table 1.** ζ-potential values and hydrodynamic size determined by DLS in water medium of the different nanosystems. Data are mean ±SD of three measurements.

| Sample | ζ-potential (mV) | Hydrodynamic size (nm) |
|---|---|---|
| MSN | −22.8 ± 0.3 | 196 ± 13 |
| MSN-G3 | +40.3 ± 1.3 | 191 ± 20 |
| MSN-G3-Ag$^+$ | +38.4 ± 0.3 | 180 ± 14 |
| MSN-G3-Zn$^{2+}$ | +41.2 ± 0.5 | 187 ± 27 |

**Table 2**. Kinetic parameters of LEVO release from the different nanosystems.

| Sample | A (µg/mg) [a] | k (h$^{-1}$) [b] | R$^2$ [c] |
|---|---|---|---|
| MSN-G3-L | 23.3 ± 0.1 | 1.00 ± 0.04 | 0.990 |
| MSN-G3-L-Ag$^+$ | 49.1 ± 0.4 | 0.66 ± 0.03 | 0.997 |
| MSN-G3-L-Zn$^{2+}$ | 220.8 ± 0.8 | 0.51 ± 0.01 | 0.996 |

[a] A is maximum amount of LEVO released; [b] k is the release rate constant; [c] R is the regression coefficients of the different fits.

**FIGURES AND FIGURE CAPTIONS**

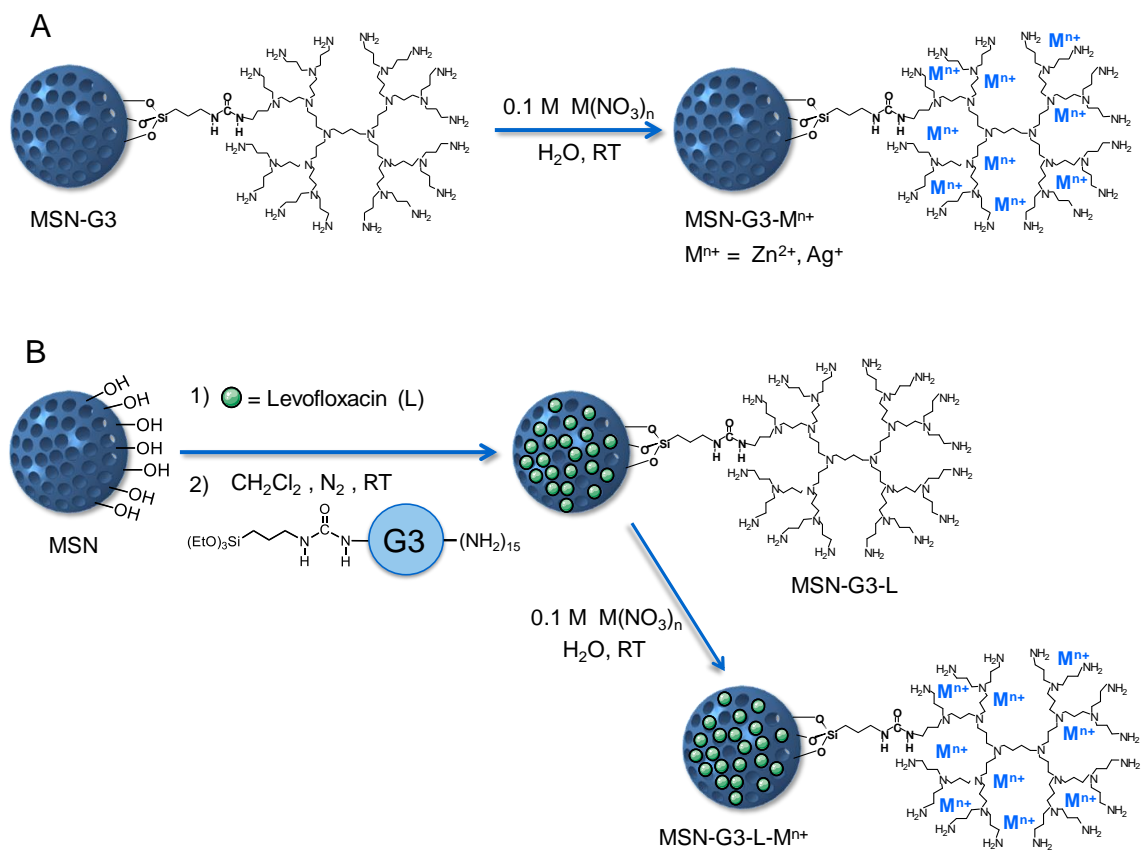

**Fig.1** Scheme of synthesis of MSN-G3-L-$M^{n+}$ materials. A) Metal cation incorporation in MSN-G3 material. B) Antibiotic loading into MSN during the functionalization with G3 dendrimer step followed by complexation of the metal cation in the dendritic wedges. L = LEVO.

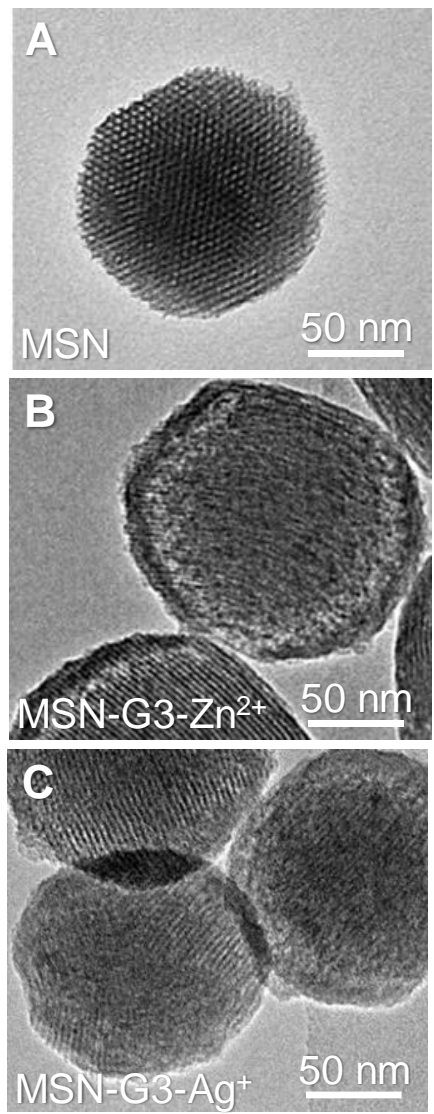

**Fig.2** TEM images corresponding to the MSN bare nanoparticles (A) and the MSN-G3-$Zn^{2+}$ (B) and MSN-G3-$Ag^+$ (C) nanosystems.

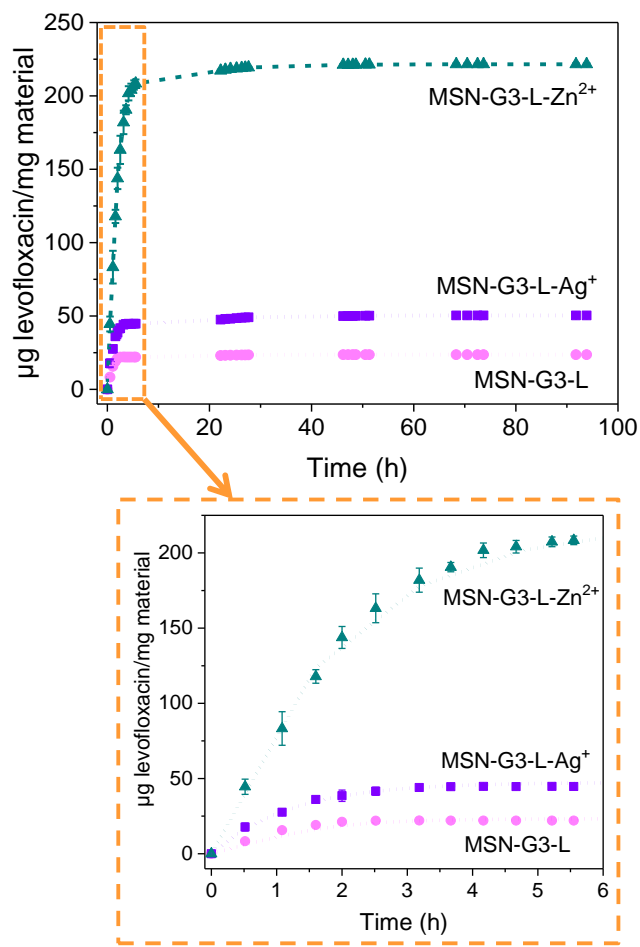

**Fig. 3** "*In vial*" cumulative LEVO release profiles from MSN-G3-L, MSN-G3-L-Ag$^+$ and MSN-G3-L-Zn$^{2+}$ samples. All data are the mean ± SD of three experiments.

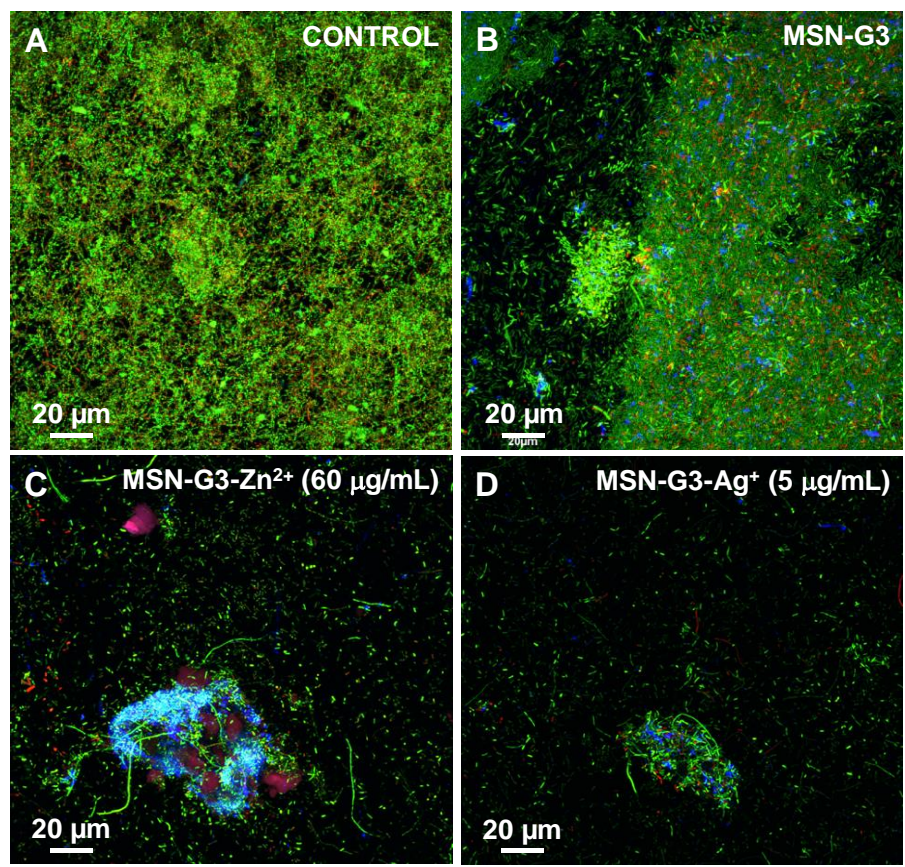

**Fig. 4** Confocal microscopy study of the antimicrobial activity of MSN-G3-M$^{n+}$ nanosystems onto Gram-negative *E. coli* biofilm. The confocal images show the biofilm preformed onto covered glass-disk after 16 h without treatment (control, **A**) and after 90 min of incubation with 5 µg/mL of MSN-G3 (**B**), 60 µg/mL of MSN-G3-Zn$^{2+}$ (**C**) and 5 µg/mL of MSN-G3-Ag$^{+}$ (**D**) nanosystems. Live bacteria are stained in green, dead bacteria in red and the protective extracellular polysaccharide matrix biofilm in blue.

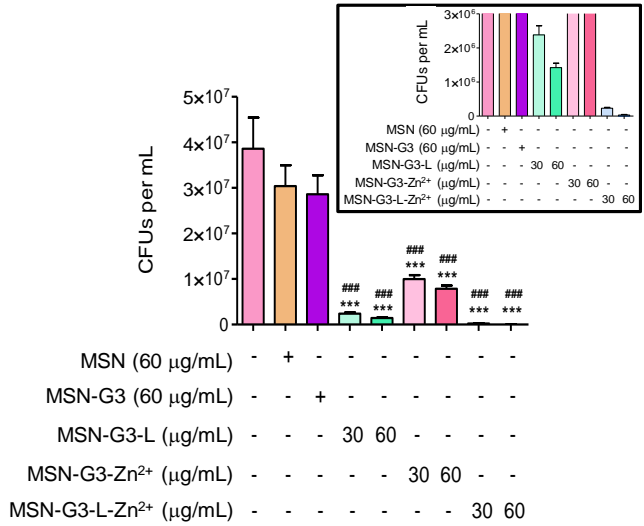

**Fig. 5 Bactericide effect MSN-G3-L-Zn$^{2+}$ nanosystem on *E. coli* biofilm.** The graphs show CFUs per mL after biofilms exposure to the nanosystems for 16 h at different concentrations, compared to biofilm without treatment as control (first bar). The inset is a magnification up to $5 \times 10^6$ CFUs/mL. Data are mean ± SD of five independent experiments. Statistical significance: *P < 0.05, compared with control; ***P < 0.001, compared with control; ###P < 0.001; compared with MSN-G3.

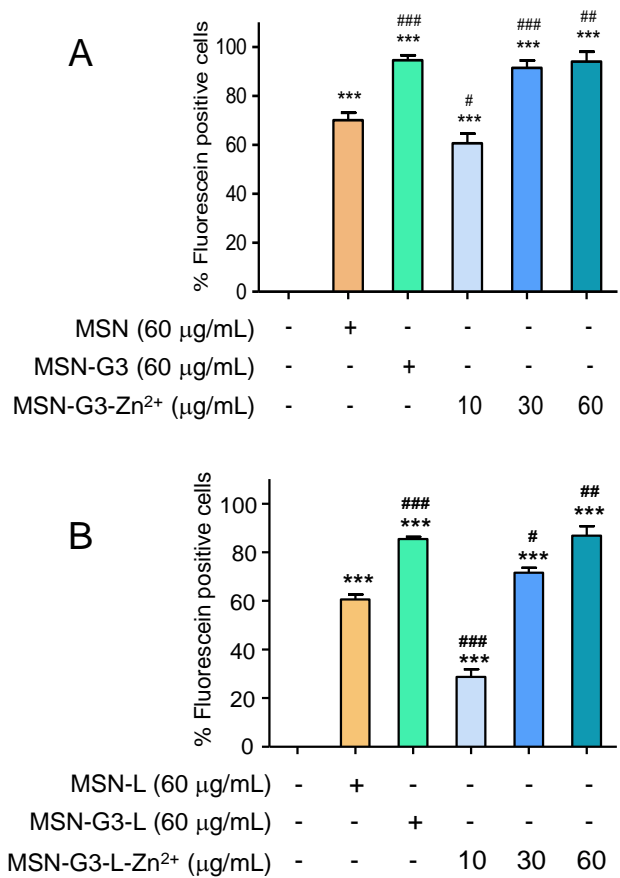

**Fig. 6** Cell uptake studies of the MSN-G3-$Zn^{2+}$ (A) and MSN-G3-L-$Zn^{2+}$ (B) nanosystems at different doses (10, 30 and 60 μg/mL) evaluated on MC3T3-E1 cells after 2 h of exposure. Percent of MC3T3-E1 cells with internalised nanoparticles was measured by flow cytometry. Control = cells cultured without materials (first bar). Data are mean ± SD of three independent experiments. Statistical significance: ***$P < 0.001$, compared with the control; #$P < 0.05$, compared with MSN; ##$P < 0.01$, compared with MSN and ###$P < 0.001$, compared with MSN.

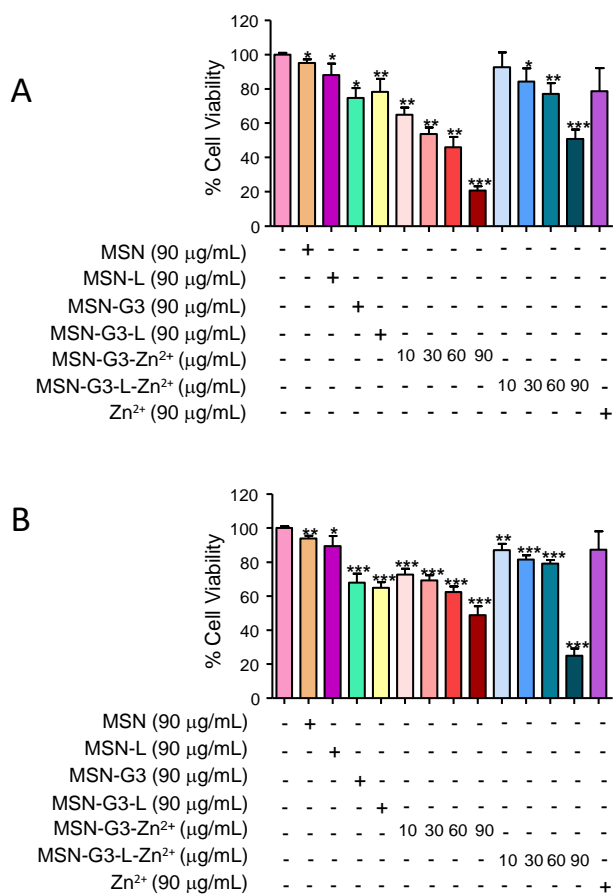

**Fig. 7** Study of cell viability of MC3T3-E1 cells exposed to different nanosystems at different concentrations. Cells were incubated with the different samples and for 2 h, and were subsequently washed and incubated in fresh media for 24 h (A) and 96 h (B) before assessing cell viability in each group. Control = cells cultured without materials (first bar). Data are means ± SD of three independent experiments. Statistical significance: *P < 0.05, compared with the control; **P < 0.01, compared with the control; ***P < 0.001, compared with the control. Materials assayed as controls (free zinc cation, MSN, MSN-G3, MSN-L and MSN-G3-L) were evaluated at the maximum doses assayed for the nanosystems which contains the metal cation and the cation plus antibiotic.

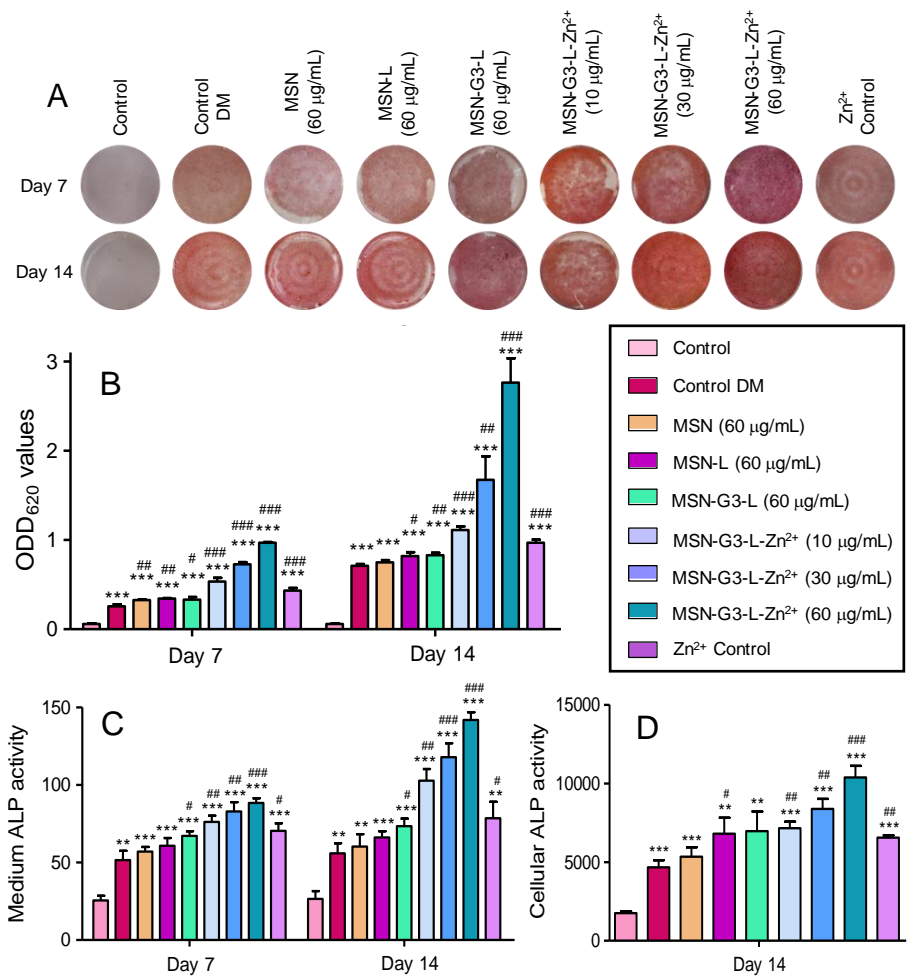

**Fig. 8** Osteoblastic differentiation assay of MC3T3-E1 cells exposed to MSN-G3-L-$Zn^{2+}$ nanosystems at different doses (10, 30 and 60 μg/mL) for 7 or 14 days in 6-well plates ($\varnothing$ = 3.5 cm; cell growth area = 9.5 $cm^2$). A control of free cation was also included ($Zn^{2+}$ Control) which consisted on the same $Zn^{2+}$ amount calculated for MSN-G3-L-$Zn^{2+}$ at 60 µg/mL. Deposits of calcium compounds stained with Alizarin Red S (A) and quantification through absorbance at 620 nm (B). Activity of ALP enzyme measured in the culture medium at days 7 and 14 (C) and within cells at day 14 (D). Data are mean ± SD of three independent experiments. Statistical significance: ***$P < 0.001$, compared with the control; #$P < 0.05$, compared with the DM control; ##$P < 0.01$, compared with the DM control and ###$P < 0.001$, compared with the DM control. DM = osteogenic or differentiation medium. Unit for medium ALP activity (nmol *p*-nitrophenol/mL/min). Unit for cellular ALP activity (nmol *p*-nitrophenol/mg protein/min).